\documentclass[traditabstract]{aa}
\usepackage{natbib}
\usepackage{txfonts}
\usepackage{graphicx}
\titlerunning{Detailed evolution of Polaris}
\authorrunning{H.~R. Neilson}

\begin{document}

\title{Revisiting the fundamental properties of Cepheid Polaris using detailed stellar evolution models}

\author{Hilding R. Neilson\inst{1}}
\institute{
   Department of Physics \& Astronomy, East Tennessee State University, Box 70652, Johnson City, TN 37614 USA
   \email{neilsonh@etsu.edu}
  }

\date{}

\abstract{
Polaris the Cepheid has been observed for centuries, presenting surprises and changing our view of Cepheids and stellar astrophysics, in general. Specifically, understanding Polaris helps anchor the Cepheid Leavitt law, but the distance must be measured precisely.  The recent debate regarding the distance to Polaris has raised questions  about its role in calibrating the Leavitt law and even its evolutionary status.  In this work, I present new stellar evolution models of Cepheids to compare with previously measured CNO abundances, period change and angular diameter.  Based on the comparison, I show that Polaris cannot be evolving along the first crossing of the Cepheid instability strip and cannot have evolved from a rapidly-rotating main sequence star.  As such, Polaris must also be at least 118~pc away and pulsates in the first overtone, disagreeing with the recent results of Turner et al. (2013).

}
\maketitle

\section{Introduction}
The North Star, Polaris (HD~8890), has fascinated people for centuries, guided explorers, and appears in the mythologies of numerous cultures.  Polaris is also the nearest classical Cepheid, making it a powerful laboratory for understanding stellar evolution as well as anchoring the calibration of the Cepheid Leavitt Law (period-luminosity relation).

However, one of the greatest hindrances to understanding Polaris is its distance.  \cite{vanLeeuwen2007} measured a distance $d = 129 \pm 2~$pc from revised Hipparcos parallaxes, whereas \cite{Turner2013} found a distance $d = 99\pm 2$~pc based on spectroscopic line ratio measurements.  The two distances are both measured precisely, yet disagree significantly.  Furthermore, these distances are particularly interesting because they suggest different pulsation properties and evolutionary histories for Polaris.

Analysis of interferometric observations suggest an angular diameter for Polaris of $\theta = 3.123 \pm 0.008$~mas \citep{Merand2007}, hence a mean radius of either $33.4 \pm 0.6~R_\odot$ or $43.5 \pm 0.8~R_\odot$ depending on which distance one considers.  If the radius is the former value then the period-radius relation \citep[e.g.][]{Gieren1997, Neilson2010, Storm2011} implies that Polaris is pulsating as a fundamental mode Cepheid, whereas if the radius is the larger value then Polaris must be pulsating as a first overtone Cepheid.  Understanding the pulsation mode is necessary for calibrating the Cepheid Leavitt Law \citep{vanLeeuwen2007} and constraining the structure of the Cepheid instability strip on the Hertzsprung-Russell diagram.

The distance debate also presents challenges for understanding the evolution of Polaris.  Assuming Polaris is at the closer distance, \cite{Turner2013} argued that Polaris is evolving along the first crossing of the Cepheid instability strip because of its large rate of period change \citep{Turner2006, Neilson2012a}.  In that case, Polaris has evolved from a main sequence star but has not yet become a red giant star.   \cite{Turner2013} also suggested that Polaris must have rotated rapidly as a main sequence star to explain the measured nitrogen and carbon abundances \citep{Usenko2005}.  On the other hand, \cite{Neilson2012a} suggested that Polaris cannot be evolving along the first crossing if it is at the farther distance because the measured rate of period change is too small relative to a first-crossing Cepheid with a similar luminosity. Hence, Polaris is evolving along the third crossing of the instability strip on the Cepheid blue loop and its observed CNO abundances are consistent with dredge up during the previous red giant stage. \cite{Neilson2012a} further suggested that to account for discrepancy between measured and predicted rates of period change, Polaris must be losing mass in an enhanced stellar wind with $\dot{M} = 10^{-7}$ -- $10^{-6}~M_\odot$~yr$^{-1}$.

Our understanding of Polaris is biased by the assumed distance, therefore measuring a precise and consistent distance is crucial.  However, there is no a priori reason to prefer one measured distance over another.  \cite{Turner2013} argued that Hipparcos parallax must be wrong because it  is inconsistent with other measurements such as space velocities of nearby stars, and main sequence fitting of Polaris's F3~V binary companion.  \cite{vanLeeuwen2013} countered that the first argument could not be reproduced, suggesting that Polaris is not a member of the local cluster. Furthermore, the F3~V stars have a large enough color distribution that would be consistent with both distances.  

\begin{figure*}[th]
\begin{center}
\includegraphics[width=0.48\textwidth]{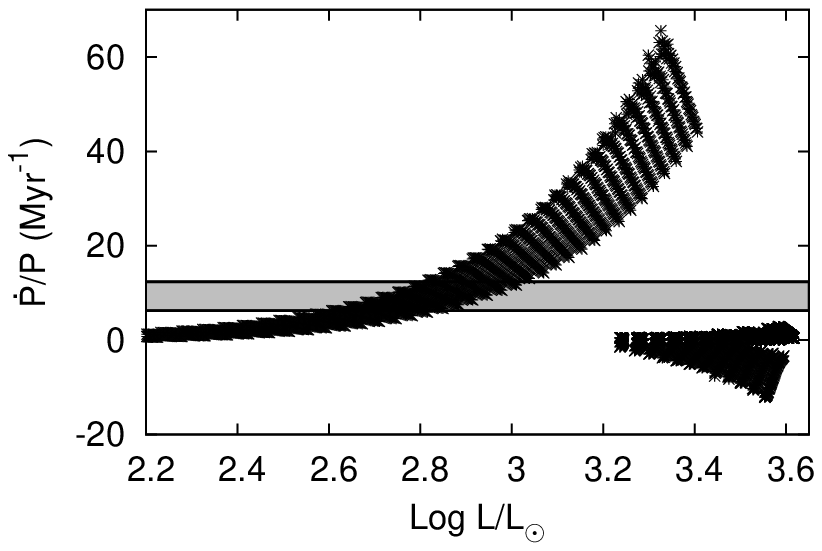}\includegraphics[width=0.48\textwidth]{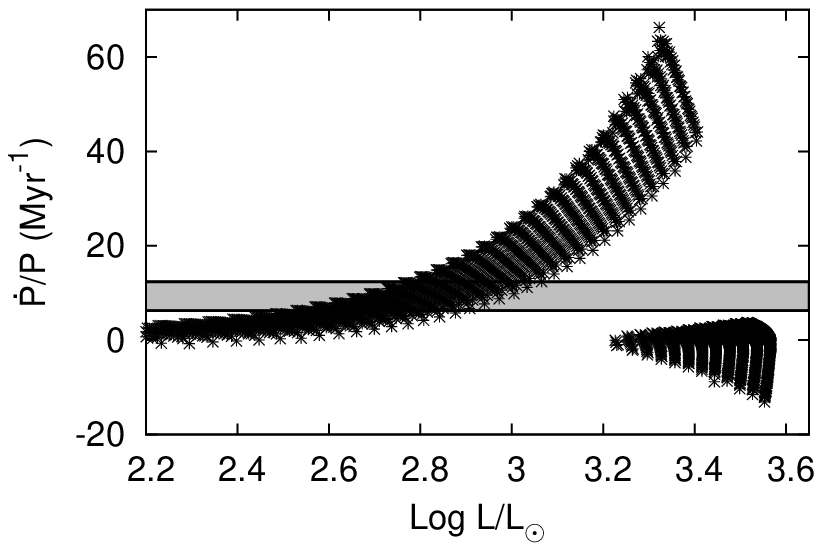}
\end{center}
\caption{Predicted rates of period change for stars crossing the Cepheid instability strip assuming different prescriptions for Cepheid mass loss: $\dot{M} = 10^{-9}~M_\odot~$yr$^{-1}$ (left) and $10^{-6}~M_\odot~$yr$^{-1}$ (right), the results for the other two cases are nearly indistinguishable.  The grey band refers to the measured rate of period change for Polaris \citep{Neilson2012a}.}
\label{plot_pdot}
\end{figure*}

There are also questions regarding the spectroscopic method employed by \cite{Turner2013}. That method is based on the FGK supergiant calibration of \cite{Kovtyukh2007} and \cite{Kovtyukh2010}, hence, stellar effective temperatures and luminosities are determined by measuring various iron line ratios from optical spectra.  However, Cepheid variable stars are not static yellow supergiants. For instance, \cite{Kervella2006} and \cite{Merand2006, Merand2007} detected infrared excesses about a sample of Galactic Cepheids, but not for the yellow supergiant star $\alpha$ Persei. There are also observations of dynamic motions in the photosphere driven by pulsation that do not occur in static yellow supergiant stars \citep[e.g.][]{Nardetto2008}. \cite{Engle2012} reported X-ray and UV observations of the nearest Cepheids including Polaris, suggesting the presence of hot ($\approx 100,000$~K) plasma in the photosphere, which may affect the ionization rates of various species.   These differences could affect the systematic uncertainties of the effective temperature and luminosity measured for Polaris.  However, in spite of these differences, \cite{Turner2013} reported a mean effective temperature $\langle T_{\rm{eff}} \rangle = 6025\pm 1$~K and $\langle M_V \rangle = 3.07\pm 0.01$, where the presented error is statistical only.  There are no systemic uncertainties presented.  Convection alone would lead to a greater variation of temperature. However, to be consistent with the Hipparcos parallax the uncertainty of the luminosity must be $\ge 0.6$ dex.

The current distance debate to Polaris is confusing our understanding of the star from the context of stellar evolution and pulsation.  In this article, I compute new state-of-the-art stellar evolution models to compare with various observations of Polaris, such as the rate of period change \citep{Turner2005, Neilson2012a}, angular diameter \citep{Merand2006}, and CNO abundances \citep{Usenko2005}.  We will also compare these models to the estimates of the distance and effective temperature, hence radius and luminosity.  Comparing stellar evolution models to the $[N/H]$ and $[C/H]$ abundances test the hypothesis that Polaris is evolving along the first crossing of the instability strip, i.e., Polaris was once a rapidly rotating during main sequence star.   Comparing models with the observed rate of period change provides insight into the mass-loss rate and which crossing of the instability strip Polaris is evolving, while comparing the models to the angular diameter constrains the radius and luminosity of the star, hence its distance.

In Sect.~2, I briefly describe the stellar evolution models computed. In Sect.~3, I describe how Cepheid rates of period change are computed from stellar evolution models. I also compute rates of period change as a function of stellar mass-loss rate to compare to the observed rate of period change.  I compare stellar evolution models with the observed nitrogen and carbon abundance to constrain the hypothesis that Polaris is a first-crossing Cepheid in Sect.~4.  In Sect.~5, I discuss the implications of these tests for the distance and pulsation mode.

\section{Methods}
I computed stellar evolution models using the \cite{Yoon2005} code.  This code was used in previous articles that include descriptions for convective core overshoot and mass loss \citep[see][]{Neilson2011, Neilson2012a, Neilson2012b}.  Previous works included overshooting and Cepheid mass loss, and in this work I consider the role of rotation to explore how rotational mixing enhances the nitrogen abundance in Cepheid models.  

To that end, I computed new stellar evolution models for masses ranging from $M = 3$ -- $6.5~M_\odot$, consistent with the astrometric-measured mass $M = 4.5^{+2.2}_{-1.4}~M_\odot$ \citep{Evans2008}.  I assumed standard solar composition \citep{Grevesse1998} and moderate overshooting $\alpha_c = 0.2$. Assuming this composition instead of the \cite{Asplund2009} composition does not significantly affect the results of this work.  The overshooting parameter was chosen to be consistent with measurements of the Large Magellanic Cloud Cepheid OGLE-LMC-CEP0227 \citep{Piet2010, Cassisi2011, Neilson2012c, Prada2012, Marconi2013}.  Different values of the overshooting parameter were considered by \cite{Neilson2012a}, who found that the rate of period change is insensitive to the value of $\alpha_c$.

I computed seven different model evolution grids.  Three grids assume negligible mass loss during the Cepheid stage of evolution, but initial rotation rates: $v_{\rm{rot}} = 50,~100$, and $200~$km~s$^{-1}$.  The remaining four grids assume negligible initial rotation but enhanced mass loss on the Cepheid instability strip such that $\dot{M}  = 10^{-9},10^{-8},10^{-7}$, and $10^{-6}~M_\odot~$yr$^{-1}$.  From these stellar evolution models, I calculated rates of period change as well as nitrogen and carbon abundances during Cepheid evolution to compare to the observations \citep{Neilson2012a, Usenko2005}.

\section{The role of mass loss}
There is growing evidence that large mass-loss rates are an ubiquitous property of Cepheids and changes the predicted rates of period change via the period-mean density relation \citep{Neilson2012b}.  Following \cite{Neilson2012a, Neilson2012b} the rate of period change is 
\begin{equation}
\frac{ \dot{P}}{P} = \frac{6}{7}\frac{\dot{L}}{L} - \frac{24}{7}\frac{\dot{T}_{\rm{eff}}}{T_{\rm{eff}}}.
\end{equation}
This relation differs from the more recent derivation by \cite{Turner2013}, who assumed a mass-period relation to derive the rate of period change.  A period-mass relation is equivalent to assuming a mass-luminosity relation for Cepheids, hence pre-supposes information regarding Cepheid evolution.  More specifically, using a period-mass relation also assumes that the Cepheid instability strip is infinitesimally narrow, i.e., ignores the width of instability strip and the primary direction of Cepheid evolution.  Because of these inconsistencies, I continued to employ the earlier derivation \citep{Neilson2012a, Neilson2012b}, although differences between this relation and the \cite{Turner2013} relation are small.

I computed predicted rates of period change for the four grids of stellar evolution models that assume negligible rotation and mass-loss rates $\dot{M} = 10^{-9}, 10^{-8}, 10^{-7}$, and $10^{-6}~M_\odot~$yr$^{-1}$. These rates of period change are shown in Fig.~\ref{plot_pdot}, as a function of stellar luminosity. The grey bar represents the measured rate of period change, $\dot{P} = 4.47 \pm 1.46~$s~yr$^{-1}$ or $\dot{P}/{P} = 9.31\pm 2.04~$Myr$^{-1}$. \cite{Neilson2012a} hypothesized that increasing the mass-loss rate would yield larger rates of period change during Cepheid evolution towards cooler effective temperatures. However, the results presented here appear inconsistent that that hypothesis.  It does not appear that changing the Cepheid mass-loss rate by four orders of magnitude leads to significant increases in the positive rate of period change by more than $1~$Myr$^{-1}$, particularly during the third crossing of the instability strip and suggests that Polaris appears to be a first-crossing Cepheid at the distance measured by \cite{Turner2013}.

\begin{figure}[t]
\begin{center}
\includegraphics[width=0.49\textwidth]{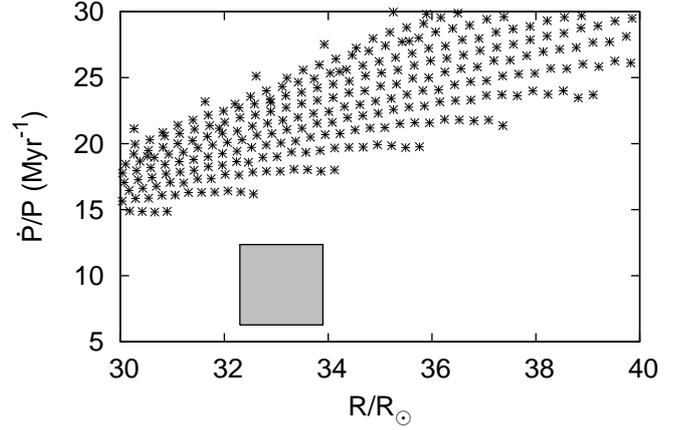}
\end{center}
\caption{Predicted rates of period change for the first crossing of the instability strip as a function of stellar radius for the computed grid of stellar evolution models assuming a Cepheid mass-loss rate of $10^{-6}~M_\odot~$yr$^{-1}$. The assumed mass-loss rate has negligible effect on the period change during the first crossing, hence the comparison using grids with other mass-loss rates has the same result.  The grey box represents the measured rate of period change and radius if the distance is $99~$pc.}
\label{plot_rpd}
\end{figure}

I explore this result by assuming the \cite{Turner2013} distance, i.e., the radius of Polaris is $R = 33.1 \pm 0.8~R_\odot$. The predicted rates of period change as a function of stellar radius are plotted in Fig.~\ref{plot_rpd} to compare with the observed period change and inferred radius given a distance, $d = 99\pm2~$pc. Cepheids on the first crossing of the instability strip do not have predicted rates of period change and radii consistent with the \cite{Turner2013} distance.   Only if the distance to Polaris is significantly less than $99~$pc might the inferred radius be consistent with measured rate of period change.  Because the time scale for stellar evolution along the first crossing is set by  dynamical time scales then the rate of period change is unlikely to vary significantly due to different assumed physics in the models.  Hence, either Polaris is much closer than measured by \cite{Turner2013} and \cite{vanLeeuwen2007}, or conversely, Polaris is accreting mass from a companion to decrease the rates of period change, or most likely, Polaris is not a first-crossing Cepheid.

\begin{figure*}[t]
\label{plot_comp}
\begin{center}
\includegraphics[width=0.49\textwidth]{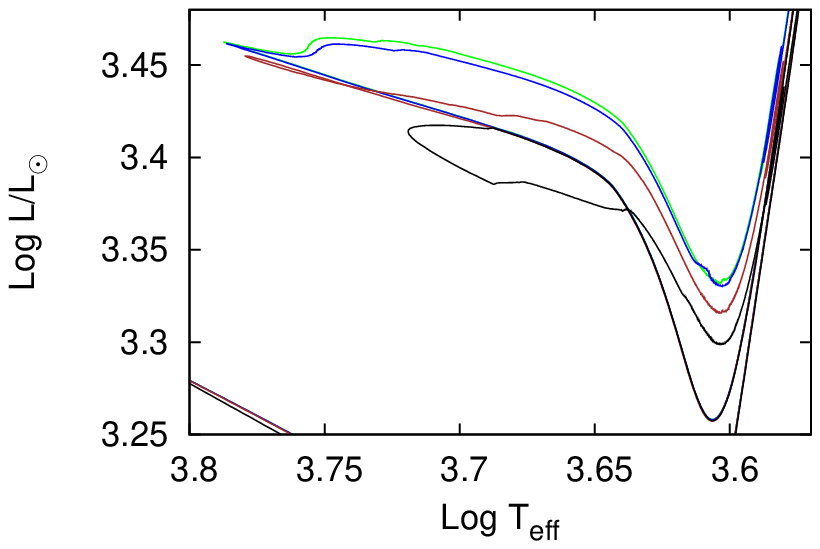}\includegraphics[width=0.49\textwidth]{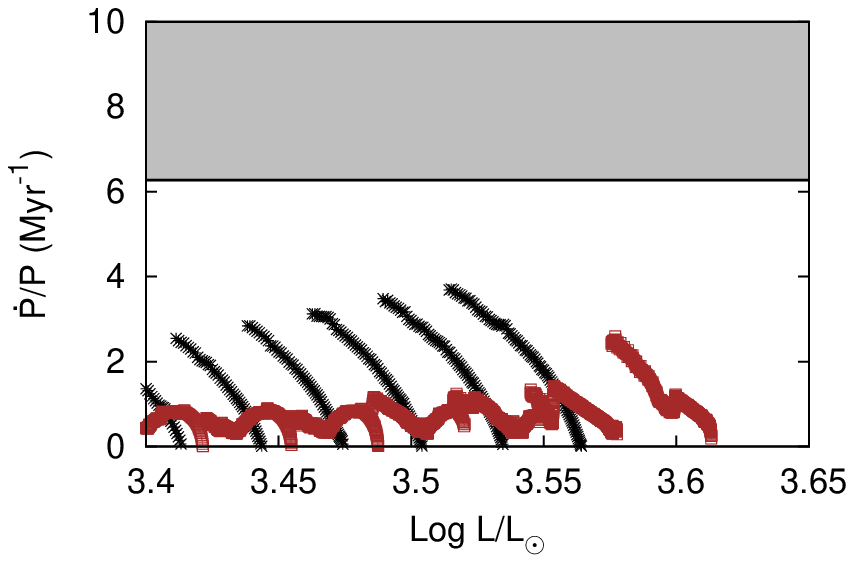}
\end{center}
\caption{(Left) Hertzsprung-Russell diagram showing the blue loop evolution of a $6~M_\odot$ star assuming different mass-loss prescriptions where $\dot{M} = 10^{-9}$ (green), $10^{-8}$ (blue), $10^{-7}$ (brown), and $10^{-6}~M_\odot~$yr$^{-1}$ (black). (Right) Predicted rates of period change as a function of luminosity for stellar evolution models with $\dot{M} = 10^{-7}$ (brown squares), and $10^{-6}~M_\odot~$yr$^{-1}$ (black stars). }
\end{figure*}

The results presented in Figs.~\ref{plot_pdot} and \ref{plot_rpd} suggest that Polaris is not a first-crossing Cepheid at either measured distance.  However, the same results also appear to contradict the hypothesis that Polaris is a third-crossing Cepheid.  \cite{Neilson2012a} suggested that the rate of period change is linearly proportional to the mass-loss rate. Yet further analysis finds that mass loss also changes the time scales for Cepheid blue loop evolution \citep{Neilson2011, Matthews2012}, as can be seen in Fig.~\ref{plot_comp}, where I plot stellar evolution tracks for $6~M_\odot$ models assuming different Cepheid mass-loss rates.  The blue loops do not differ significantly until one assumes mass-loss rates of the order $10^{-6}~M_\odot~$yr$^{-1}$. The hottest effective temperature for the tip of the blue loop is about $T_{\rm{eff}} = 6150~$K for smaller mass-loss models whereas the blue loop tip occurs at $T_{\rm{eff}} = 5370~$K when $\dot{M} = 10^{-6}~M_\odot~$yr$^{-1}$, a difference $\Delta T_{\rm{eff}}/ T_{\rm{eff}} \approx 15\%$.  If one assumes the blue loop lifetime for the $10^{-6}~M_\odot~$yr$^{-1}$ models, but with the blue loop width for models with smaller mass-loss rates, then the predicted rates of period change would increase by more than 50\%. Under that assumption it is possible that the period change for Polaris is consistent with a mass-loss rate of the order $10^{-6}~M_\odot~$yr$^{-1}$.

This argument is speculative because the physics that drives blue loop evolution is still uncertain. Cepheid mass loss is one ingredient that appears to shorten blue loops \citep{Neilson2011, Matthews2012}, but metallicity and convective core overshooting play roles \cite[e.g.][]{Cassisi2011, Neilson2012c}. \cite{Kippenhahntext} noted that the blue loop is also a function of the relative size of the helium core and the hydrogen burning shell, hence any changes in this ratio dramatically changes the structure of the blue loop.  As such, Polaris is most likely a third-crossing Cepheid and cannot be a first-crossing Cepheid as the rate of period change along that crossing is described solely by dynamic time scales.  By all measures \citep{vanLeeuwen2007, Turner2013}, Polaris is too distant to be consistent with first-crossing stellar evolution.

\section{The role of rotation}
While the period change implies that Polaris is not a first-crossing Cepheid, it is not conclusive. A second method to constrain the evolutionary stage of Polaris is its nitrogen and carbon abundances. \cite{Usenko2005} measured abundances, $[N/H] = 0.4$, $[C/H] = -0.17]$, and $[O/H] = 0$. Nitrogen enhancement and carbon diminishment have two possible causes: dredge up when a star enters the red giant branch or rotational mixing during main sequence evolution. If only the former process is important then any Cepheid with enhanced nitrogen abundance must be evolving along the blue loop. Therefore, we test whether rotational mixing is important by computing grids of stellar evolution models with different initial rotational velocity and explore the predicted nitrogen enhancement and rotation rates for Cepheids evolving along the first crossing of the instability strip.

\begin{figure*}[t]
\label{rotation}
\begin{center}
\includegraphics[width =0.49\textwidth]{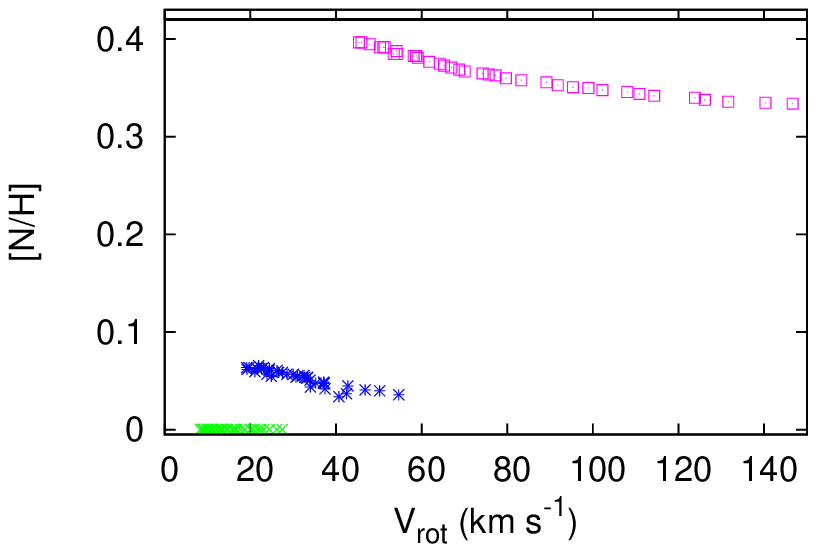}\includegraphics[width =0.49\textwidth]{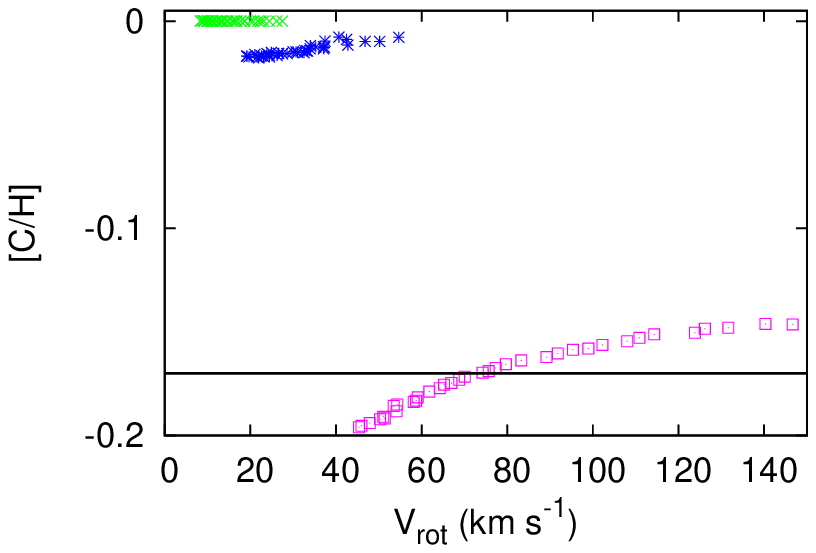}
\end{center}
\caption{Predicted rotation rates and nitrogen (left) and carbon (right) abundance for Cepheids crossing the instability strip for the first time.  Predictions are based on three initial rotation rates: $v = 50$ (green x's), 100 (blue stars), and $200~$km~s$^{-1}$ (magenta squares), while the horizontal line represents the abundance measured by \cite{Usenko2005}.}
\end{figure*}
I plot predicted nitrogen and carbon abundances as a function of rotation rate for computed first-crossing Cepheids in Fig.~\ref{rotation} for the three cases with initial velocity, $v_{\rm{init}} = 50, 100,$ and $200~$km~s$^{-1}$. The nitrogen abundances are enhanced and carbon abundances are depressed just enough to be consistent with observations only when $v_{\rm{init}} = 200~$km~s$^{-1}$. It should be noted that \cite{Usenko2005} presented the measured nitrogen abundance without uncertainties, hence it is difficult to make a quantitative comparison.

A more striking result is the predicted Cepheid rotation rates. \cite{Bersier1996} measured rotation rates for a sample of Galactic Cepheids to be of the order $10~$km~s$^{-1}$, where the most rapidly-rotating Cepheids are first-crossing Cepheids. The model Cepheids have not lost enough angular momentum to be consistent with the observed slow rotation rates. Also, the loss of angular momentum during main sequence evolution inhibits rotational mixing, hence it is difficult to envision a scenario in which rotational mixing is still efficient enough while sufficient angular momentum is lost to be consistent with observations. Furthermore, the predicted rotation rates are a function of stellar mass, i.e., the most massive Cepheids are the slowest rotators. Hence, to be consistent with a first-crossing Cepheid, Polaris would have to have mass $M > 6~M_\odot$, but that mass is inconsistent with pulsation calculations \citep{Caputo2005}.  The results presented in Fig.~\ref{rotation} contradict the hypothesis that Polaris is a first-crossing Cepheid. 

The nitrogen and carbon abundances measured by \cite{Usenko2005} are consistent with both a rotationally-mixed first-crossing Cepheid and a Cepheid evolving along the blue loop.  However, the models with large initial velocities still rotate at speeds greater than observational limits, suggesting that either the main sequence progenitor of Polaris was a slow rotator or angular momentum loss during main sequence evolution must be more efficient. This implies that the surface abundances in Polaris were not enhanced by rotational mixing.

The only potential case where the Polaris could be a first-crossing Cepheid with both enhanced nitrogen abundance and negligible rotation is if Polaris evolved from a small population of non-rotating nitrogen-rich massive main sequence stars \citep{Hunter2008, Brott2011}. However, this population is known only in the LMC and has not yet been detected in the Milky Way. Also, stars in this population have masses $> 10~M_\odot$ making it a highly unlikely explanation for the evolution of Polaris.

\section{Constraining the distance}
\begin{figure}[t]
\begin{center}
\includegraphics[width=0.49\textwidth]{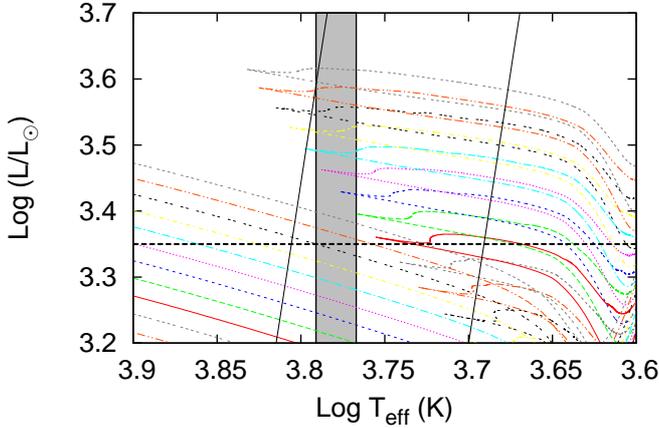}
\end{center}
\caption{HR diagram showing stellar evolution tracks with zero initial rotation and mass-loss rate, $\dot{M} = 10^{-9}~M_\odot$~yr$^{-1}$,  focused on the region of the diagram consistent with the effective temperature of Polaris, $T_{\rm{eff}} = 6015\pm 170~$K \citep{Usenko2005}, denoted by the grey box.  The diagonal black lines form the boundaries of the Cepheid instability strip, while the horizontal line denotes the likely minimum blue loop luminosity consistent with the effective temperature of Polaris.  The brightest blue loop represents an initial stellar mass of $6.5~M_\odot$ (grey dashed), while the dimmest blue loop shown is for a $5.4~M_\odot$ model (black dashes).  Each model differs by $\Delta M = 0.1~M_\odot$.} 
\label{hrs}
\end{figure}

The debate over the distance to Polaris affects arguments regarding its pulsation mode and which crossing of the instability strip it is evolving. I have shown that Polaris cannot be evolving along the first crossing of the instability strip but instead the third crossing. Based on this result and to constrain the minimum distance, using the predicted stellar evolution models, I measured the minimum luminosity for Polaris as a Cepheid evolving along the blue loop, as shown in Fig.~\ref{hrs}.

 The minimum luminosity for Polaris, estimated from Fig~\ref{hrs}, is about $\log (L/L_\odot) = 3.35$.  This luminosity is smaller than the minimum luminosity for a stellar evolution model evolveing along the blue loop that is also consistent with the measured effective temperature. Because the underlying physics driving blue loop stellar evolution is still uncertain, the tip of the blue loop can shift by about 100 - 200~K.  Therefore, the smaller value for the luminosity is conservative. The minimum distance to Polaris is inferred by using this minimum luminosity, the maximum effective temperature, $T_{\rm{eff, max}} = 6185~$K \citep{Usenko2005}, and the measured angular diameter \citep{Merand2007}.  Using these values, the minimum distance to Polaris is $d \ge 118~$pc. 

This inferred minimum distance is significantly farther than the distance measured by \cite{Turner2013} and assumes the effective temperature is about $150~$K greater than the mean effective temperature also measured by \cite{Turner2013}.  Therefore, it can be concluded that Polaris must be more distant and is consistent with the previously measured Hipparcos parallax.  Similarly, the pulsation mode for Polaris has been previously derived based on the measured distance, hence radius, and combined with a period-radius relation \citep{Neilson2012a, Turner2013}. At a distance of 118~pc, the predicted mean radius is $R = 41.1~R_\odot$.  Using the period-radius relations from \cite{Neilson2010}, the predicted radius for Polaris is $R \approx 35$ -- $36~R_\odot$ if Polaris is a fundamental-mode Cepheid or $R \approx 44$ -- $46~R_\odot$ if Polaris is a first-overtone Cepheid.  The hypothesis that Polaris is a fundamental-mode Cepheid is inconsistent with the inferred minimum distance, but is more consistent with a first-overtone pulsation mode.

The results from stellar evolution calculations suggest that Polaris is a third-crossing Cepheid pulsating in the first overtone at a distance $> 118~$pc.  However, \cite{Turner2005} suggested that the measured proper motion and radial velocity of Polaris is inconsistent with other stars in the vicinity of Polaris at the Hipparcos distance. That result challenges our understanding of the properties of Polaris, but  can be explained by Polaris being a runaway star.  There are other examples of Cepheid runaway stars, such as the prototype $\delta$ Cephei \citep{Marengo2010} and potentially, GU Nor \citep{Majaess2013}.

Another possible insight into the evolutionary status of, and distance to, Polaris is its apparent brightening over the past century.  \cite{Engle2014} showed that Polaris appears to be about 0.05 $V$-band magnitude brighter now than a century ago.  If confirmed, this suggests, again, that Polaris is not evolving on the first crossing of the instability strip since a star evolving from the end of the main sequence to the red giant branch has a decreasing luminosity.  All models computed in this work display that property during the first crossing of the instability strip, regardless of mass loss and rotation.  On the other hand, the luminosity increase during blue loop stellar evolution is at least $10^{-4}~L_\odot$~per century, orders of magnitude smaller than that measured by \cite{Engle2014}, hence raising more questions.

\section{Conclusions}
The debate over the distance to Polaris is important for the calibration of the Cepheid Leavitt Law, as the nearest Cepheids will calibrate the zero-point \citep[e.g.][]{Freedman2012}. Resolving the debate is also important for constraining stellar evolution physics, and constraining the properties of other Cepheids, which will become particularly important as GAIA observations will measure parallaxes for thousands of Cepheids \citep{Windmark2011}. 

In this work, I presented new stellar evolution models to reanalyze the fundamental properties of Polaris, this time without assuming a distance.  The observed rate of period change is found to be inconsistent with stellar evolution models for Cepheids, for both the first and third crossings of the instability strip, regardless of whether the \cite{vanLeeuwen2013} or \cite{Turner2013} distance is assumed.  There is no obvious theory where Polaris is consistent with stellar evolution on the first crossing of the instability strip, since that time scale is determined by dynamical time scales.   On the other hand, mass loss could explain the rate of period change for a Cepheid on the third crossing, even though mass loss also reduces the effective temperature width of the Cepheid blue loop making it difficult to compare with the observed properties of Polaris.  However, it can be noted that the measured period change is inconsistent with a first-crossing Cepheid.

I further computed new rotating stellar evolution models to explore the hypothesis that Polaris is a first-crossing Cepheid with nitrogen and carbon abundances changed by rotational mixing.  Rotational mixing models predict abundances roughly consistent with measurements by \cite{Usenko2005} during the first-crossing of the instability strip, however, the predicted rotation rates are too large to be consistent with the measured turbulent velocity for Polaris \citep{Usenko2005} and rotational velocities for a sample of Galactic Cepheids \citep{Bersier1996}. As such, the properties of Polaris are inconsistent with being a first-crossing Cepheid and must be a third-crossing Cepheid. Because Polaris is a third-crossing Cepheid, then it must also be at a distance greater than $118$~pc, significantly greater than that measured by \cite{Turner2013} and must also be pulsating as a first-overtone Cepheid.  While the results of \cite{Turner2013} are inconsistent with stellar evolution calculations, there is still much to explore about the nearest Cepheid, Polaris to constrain the physics behind stellar multiplicity \citep[e.g.][]{Evans2013}, mass loss \citep{Neilson2012a, Neilson2012b}, ultraviolet and x-ray radiation \citep{Engle2012}, the apparent brightening of Polaris \citep{Engle2014} and late-stage evolution \citep[e.g.][]{Langer2012} as we approach an era of precision stellar astrophysics. 
\acknowledgements

This work has been supported by the NSF grant (AST-0807664).

\bibliographystyle{aa} 

\bibliography{pol} 

\end{document}